\documentclass[letterpaper,conference]{IEEEtran}

\usepackage{cite}
\usepackage{graphicx}
\usepackage{subcaption}
\usepackage[cmex10]{amsmath}
\usepackage{algpseudocode}
\usepackage{algorithmicx}
\usepackage{array}
\usepackage{textcomp}
\usepackage{url}
\usepackage{bbm}
\usepackage[T1]{fontenc}
\usepackage{multirow}
\usepackage{epsfig}
\usepackage{amsfonts,amssymb,multirow,bigstrut,booktabs,ctable,latexsym}
\usepackage{mathtools}
\usepackage[mathscr]{eucal}
\usepackage{verbatim}

\usepackage{amsthm}
\usepackage{algorithm}
\usepackage[normalem]{ulem}

\IEEEoverridecommandlockouts
\def\BibTeX{{\rm B\kern-.05em{\sc i\kern-.025em b}\kern-.08em
		T\kern-.1667em\lower.7ex\hbox{E}\kern-.125emX}}
\begin{document}

	\newtheorem{definition}{Definition}
	\newtheorem{lemma}{Lemma}
	\newtheorem{theorem}{Theorem}
	\newtheorem{example}{Example}
	\newtheorem{proposition}{Proposition}
	\newtheorem{remark}{Remark}
	\newtheorem{assumption}{Assumption}
	\newtheorem{corrolary}{Corrolary}
	\newtheorem{property}{Property}
	\newtheorem{problem}{Problem}
	\newcommand{\argmin}{\arg\!\min}
	\newcommand{\argmax}{\arg\!\max}
	\newcommand{\st}{\text{s.t.}}
	\newcommand \dd[1]  { \,\textrm d{#1}  }
    \newtheorem{excont}{Example}
    \renewcommand{\theexcont}{\theexample}

		
 \title{Who is Responsible? Explaining Safety Violations in Multi-Agent Cyber-Physical Systems}

 	\author{Luyao Niu$^{1}$, Hongchao Zhang$^{2}$, Dinuka Sahabandu$^{1}$,\\ Bhaskar Ramasubramanian$^{3}$, Andrew Clark$^{2}$, Radha Poovendran$^{1}$%
		\thanks{$^{1}$Department of Electrical and Computer Engineering, 
			University of Washington, Seattle, WA, USA. 
			{\tt\small \{luyaoniu, sdinuka, rp3\}@uw.edu}}
   \thanks{$^2$Department of Electrical and Systems Engineering, Washington University in St. Louis, St. Louis, MO, USA.
			{\tt \{hongchao, andrewclark\}@wustl.edu}}
\thanks{$^3$Electrical and Computer Engineering, Western Washington University, Bellingham, WA, USA.
			{\tt ramasub@wwu.edu}}
\thanks{This work was supported by the Office of Naval Research grant N0014-23-1-2386, National
Science Foundation grants CNS 1941670, CNS 2153136, and Air Force Office of Scientific Research
grants FA9550-22-1-0054 and FA9550-23-1-0208.}
	}

	\maketitle
	
	\begin{abstract}
Multi-agent cyber-physical systems are present in a variety of applications. {Agent decision-making can be affected due to errors induced by uncertain, dynamic operating environments or due to incorrect actions taken by an agent. When an erroneous decision that leads to a violation of safety is identified, assigning responsibility to individual agents is a key step towards preventing future accidents.}
Current approaches to carrying out such investigations require human labor or high degree of familiarity with operating environments. 
Automated strategies to assign responsibility can achieve significant reduction in human effort and associated cognitive burden.  

In this paper, we develop an automated procedure to assign responsibility for safety violations to actions of any single agent in a principled manner. 
We ground our approach on reasoning about safety violations in road safety. 
When provided with an instance of a safety violation, we use counterfactual reasoning to create alternate scenarios that determine how different outcomes might have been achieved if a specific action or set of actions was replaced by another action or set of actions. 
We devise a metric called the \emph{degree of responsibility (DoR)} for each agent. 
The DoR uses the Shapley value to quantify the relative contribution of each agent to the observed safety violation, thus serving as a basis to explain and justify future decisions.
We devise both heuristic techniques and methods based on the structure of agent interactions to improve scalability of our solution as the number of agents increases. 
We consider three instances of safety violations from the National Highway Traffic Safety Administration (NHTSA). 
We carry out experiments using representations of the three scenarios using 
the CARLA urban driving simulator. 
Our results indicate that the DoR enhances explainability of decision-making and assigning accountability for actions of agents and their consequences. 
	\end{abstract}

    \section{Introduction}\label{sec:intro}

The operation of complex cyber and cyber-physical systems (CPS) such as autonomous cars relies heavily on the seamless integration of physical components with software-based decision making algorithms and procedures. 
When CPS are deployed in the real-world, over the horizon of its operation, each system will have to solve a sequential decision making problem in order to `optimally' satisfy desired objectives (e.g., safety, reachability) in uncertain environments. 
The safe operation of CPS 
is enforced as a constraint, typically on the state of the system \cite{van2008reciprocal,alonso2013optimal,borrmann2015control,chen2021scalable,wang2017safety,chen2020guaranteed}. 
Examples of safety constraints include (i) any pair of agents should be at least some distance $d$ apart in order to avoid collisions, or (ii) no agent should enter a specified `forbidden' region in the environment. 

There is a large body of work that has developed solutions with guarantees of safety during the design or operation phases of multi-agent CPS \cite{van2008reciprocal,alonso2013optimal,borrmann2015control,chen2021scalable,wang2017safety,chen2020guaranteed,yel2020assured,lavaei2023compositional,chu2020multi}. 
Two well-established directions of research in this domain are verification during design \cite{yel2020assured,lavaei2023compositional,narasimhan2023safe} and synthesizing safety-critical controllers during deployment \cite{van2008reciprocal,alonso2013optimal,borrmann2015control,chen2021scalable,wang2017safety,chen2020guaranteed}. 
However, uncertainties or changes in the operating environment or erroneous agent behavior might result in safety violations. 
For example, a controller synthesized to be safe for a car driven in clear weather may not be safe when it is raining \cite{cai2020real,yang2022interpretable}. 

Devising strategies to mitigate safety violations will be especially crucial for successful large-scale deployment of multi-agent CPS. 
A particular multi-agent CPS that we consider is a road network with emphasis on urban driving. 
The reason for this is twofold: (i) erroneous agent decisions (i.e., incorrect actions by vehicle drivers) have been known to result in safety violations (i.e., cause an accident), and (ii) road traffic accidents have been known to be a leading cause of death and billions of dollars of economic losses in the US \cite{trafficaddicent,AVcollision}. 

When errors in decision making by agents lead to safety violations, a key challenge is to identify and assign responsibility to actions of individual agents. 
This will inform development of measures to prevent future accidents, e.g., by correcting flaws in perception/control algorithms, or introducing new safety rules. However, current approaches to carrying out such investigations are either application/instance-specific \cite{yawovi2022wrong} or require human labor \cite{winfield2021robot,omeiza2021towards}, which can incur large costs. 
In some cases, a high level of familiarity with the operating environment \cite{beck2023automated} might also be required, which can place significant cognitive burden on human analysts. 
Designing automated procedures to assign responsibility will aid in reducing the cognitive overload on human operators. 
Further, such procedures can serve as an auditable basis to explain and justify future decisions, including those that might be made by human analysts. 

In this paper, we develop foundations of an automated procedure to assign responsibility to actions of any single agent relative to other agents after a safety violation in a principled manner. 
When presented with an example 
(in our setup, a trajectory) of a safety violation, we use \emph{counterfactual reasoning} \cite{bottou2013counterfactual,epstude2008functional} to identify and explain agent roles and responsibilities. 
Counterfactual reasoning provides a way of creating alternate scenarios, i.e., \emph{counter to the facts}, that seek to determine how different an outcome might have been if a specific action or set of actions was replaced by another action or set of actions. 
By systematically comparing outcomes 
in the \emph{counterfactual worlds} with actual observed outcomes, we will be able to identify which agents were responsible for the observed safety violation, 
relative to other agents. 

We reason about responsibility of individual agents through 
a multi-agent Markov decision process (MMDP), a trajectory of which has been observed to violate safety. 
The primary challenge in this setting is that we only have access to one trajectory that violated the safety constraint, and we lack information on how agents would behave when taking some alternative actions. 
We propose to mitigate this challenge by formulating a coalitional game on the MMDP, where a coalition consists of a subset of agents who choose their actions in a counterfactual world according to a safe control policy, i.e., a policy that minimizes probability of violating the safety constraint. 
For a given coalition, we construct a utility function that captures whether agents within the coalition had alternate actions available to them that would have prevented violation of safety. 
We devise a metric that we term the \emph{degree of responsibility (DoR)}. 
The DoR leverages insights from the Shapley value \cite{roth1988shapley} from game theory to quantify the relative contribution of each agent to the observed safety violation. 
\textcolor{black}{Different from causal reasoning \cite{halpern2005causes}, our DoR is not necessarily a binary value when assigning responsibilities.}

To scale our solution to large numbers of interacting agents, we develop both heuristic solutions and methods based on the structure of agent interactions to improve efficiency of DoR computation. 
We design an algorithm to characterize the marginal contribution of an agent to the safety violation, and use this algorithm to isolate a subset of potentially responsible agents.
We also show that when the MMDP satisfies an exponential decay property \cite{qu2020scalable}, the DoR of each agent can be obtained with lower computational and memory complexities. 

We ground our approach on providing explanations for safety violations by automobiles, motivated by the economic and social impacts of road traffic accidents \cite{trafficaddicent,AVcollision}. 
We consider three instances of safety violations from the National Highway Traffic Safety Administration (NHTSA) \cite{crashscenario}. 
We carry out extensive experiments using representations of the three scenarios within  
the CARLA urban driving simulator \cite{dosovitskiy17} to identify and quantify agent responsibility for a safety violation using counterfactual reasoning. 
Our results indicate that our DoR metric enhances explainability of decision-making and assigning accountability for actions of agents and their consequences. 
Moreover, comparing the DoRs of agents yields an ordered list of responsibilities, which can serve as a sound basis to explain and justify subsequent decisions, including those potentially taken by humans. 
\textcolor{black}{In these scenarios, the actions taken by agents are coupled and their epistemic states are often unknown, making existing reasoning techniques \cite{chockler2004responsibility} inadequate.}

The remainder of this paper is organized as follows: 
Sec. \ref{sec:relatedwork} discusses related work.
We formulate the problem studied in Sec. \ref{sec:formulation} and detail our solution approach in Sec. \ref{sec:solutionapproach}. 
Sec. \ref{sec:efficientcomputation} presents a scalable way to compute the $DoR$ when there are a large number of agents. 
We perform experiments to validate our approach in Sec. \ref{sec:experiments} 
and conclude the paper in Sec. \ref{sec:conclusion}. 
 





    \section{Related Work}\label{sec:relatedwork}

Guaranteeing safety in multi-agent systems (MAS) has been extensively studied. 
Typical solutions include control-theoretic based approaches and learning-based approaches.
When the agents follow known dynamics or incur bounded uncertainty in the worst-case, control-theoretic approaches can be applied to synthesize controllers for the agents \cite{van2008reciprocal,alonso2013optimal,borrmann2015control,chen2021scalable,wang2017safety,chen2020guaranteed}. 
When the dynamics followed by the agents are unknown, learning-based approaches \cite{shalev2016safe,cheng2020safe,chen2017socially,lowe2017multi,everett2018motion,zhang2018fully,qin2021learning,Aly2021Safe,lavaei2023compositional} can be utilized to learn the controllers with safety guarantees.
The aforementioned studies mainly focus on designing controllers with safety guarantees for MAS or verifying safety.
In this paper, we consider the scenarios where safety is observed to be violated by the multi-agent system.
Our goal is to quantify the individual agent's responsibility for the observed safety violation.

Although there have been extensive studies on guaranteeing safety of multi-agent CPS, safety may still be violated in practical applications due to uncertainties, faults, and errors.
There are two major categories of approaches to investigate safety violations.
The first category utilizes data records on safety violations to identify key risk factors.
For example, some factors such as weather, lighting, and roadway surfaces 
are identified in \cite{torres2021investigating}. 
In \cite{petrovic2020traffic}, statistical analysis is performed to analyze autonomous vehicle collisions, with particular emphasis on collision types and maneuvers as well as errors of drivers of conventional vehicles.
Crash narratives are utilized in \cite{lee2023advancing} to identify the factors contributing to autonomous vehicle crashes.
The second category focuses on assisting the accident investigators to explain the safety violations.
In \cite{winfield2021robot}, a why-because list is constructed by the investigation team to explain safety violations of social robots.
In \cite{omeiza2021towards}, the authors developed a tree-based representation to explain the behaviors of autonomous vehicles. 
A data pipeline is proposed in \cite{beck2023automated} for investigators of autonomous vehicle crashes to reconstruct the scene of accidents using raw data from sensors mounted by the vehicles.
In contrast, this paper aims to develop an automated procedure to quantify the responsibilities of agents in multi-agent CPS for safety violations, which has not been investigated in the aforementioned studies.

In \cite{yawovi2022wrong}, responsibilities for crossroad vehicle collisions are assessed based on traffic rules.
In this paper, we leverage the Shapley value \cite{roth1988shapley} to develop an explainable metric to quantify the responsibility of each individual agent for the safety violation.
The Shapley value has also been adopted in artificial intelligence to explain why machine models produce certain outputs for given inputs \cite{rozemberczki2022shapley}.
However, quantifying responsibility in multi-agent CPS studied in this paper presents unique challenges, distinct from \cite{yawovi2022wrong,rozemberczki2022shapley}.
That is, quantification of each agent's responsibility in our work requires decoupling and isolating not only effects of actions taken by other agents but also the influence of its own prior actions.


%

%

 
	\section{System Model and Problem Formulation}\label{sec:formulation}

We consider a multi-agent CPS consisting of a collection of agents, denoted $\mathcal{N}=\{1,\ldots,N\}$.
The dynamics and interactions among the agents are modeled as a multi-agent Markov decision process (MMDP), which is defined as follows.
\begin{definition}[Multi-Agent Markov Decision Process (MMDP)]\label{def:MDP}
    A multi-agent Markov decision process $\mathbb{M}$ is a tuple $\mathbb{M}= (\{\mathcal{S}_i\}_{i=1}^{N},\{\mathcal{A}_i\}_{i=1}^{N},Pr)$, where $\mathcal{S}_i$ is a finite set of states of agent $i\in\mathcal{N}$, $\mathcal{A}_i$ is a finite set of actions available to agent $i$, and $Pr: \mathcal{S}\times \mathcal{A}\rightarrow \Delta(\mathcal{S})$ is a transition function where $\mathcal{S}=\prod_{i=1}^N\mathcal{S}_i$ and $\mathcal{A}=\prod_{i=1}^N\mathcal{A}_i$ are the joint state and action spaces of all agents,  and $\Delta(\mathcal{S})$ denotes the set of probability distributions over set $\mathcal{S}$.
    The probability of transitioning from state $s=(s_1,\ldots,s_N)$ to state $s^\prime=(s_1',\ldots,s_N')$  when the agents take actions $a=(a_1,\ldots,a_N)$ is written as $Pr(s'|s,a)$.
\end{definition}

We define the \emph{joint non-stationary deterministic policy} of the agents on the MMDP $\mathbb{M}$ as $\pi:\mathcal{S}\times\mathbb{Z}\rightarrow\mathcal{A}$, which specifies a joint action $a\in\mathcal{A}$ of all agents for any  joint state $s\in\mathcal{S}$ and stage $t\in\mathbb{Z}$.
A \emph{non-stationary deterministic policy} of an agent $i$ is a mapping $\pi_i:\mathcal{S}\times\mathbb{Z}\rightarrow\mathcal{A}_i$ such that $\pi_i(s,t)\in \mathcal{A}_i$ for any joint state $s\in\mathcal{S}$ and stage $t\in\mathbb{Z}$.
Thus, for any joint state $s\in\mathcal{S}$, we can represent the joint policy $\pi(s,t)$ as a tuple $\pi(s,t)=(\pi_1(s,t),\ldots,\pi_N(s,t))$. 

Given the MMDP $\mathbb{M}$, a \emph{finite path} $\rho$ on $\mathbb{M}$ is a sequence of state-action pairs of finite length.
Given a finite path $\rho = s^0, a^0, s^1,\ldots, a^{T-2}, s^{T-1}$, the path of each individual agent $i\in\mathcal{N}$ can be recovered as $\rho_i = s_i^0, a_i^0, s_i^1,\ldots, a_i^{T-2}, s_i^{T-1}$.
In the remainder of this paper, we denote the state $s^t$ (resp. $s_i^t$) visited by path $\rho$ (resp. $\rho_i$) as $\rho^t$ (resp. $\rho_i^t$).
Given a time horizon $T$, a joint policy $\pi$, and initial state $s^0$ of the MMDP $\mathbb{M}$, it induces a collection of paths, where each path $\rho = s^0, a^0, s^1,\ldots, a^{T-2}, s^{T-1}$ satisfies $a^t=\pi(s^t,t)$ and $Pr(s^{t+1}|s^t,a^t)>0$ for all $t=0,\ldots,T-2$.

\begin{figure}[h]
    \centering
    \includegraphics[width=0.48\textwidth]{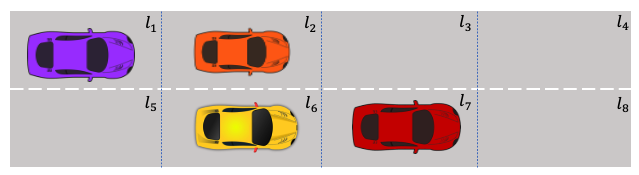}
    \caption{This figure presents a road segment consisting of two lanes. The road segment is discretized into eight discrete locations. Four vehicles share the road segment. }
    \label{fig:example}
\end{figure}

We illustrate the MMDP in Definition \ref{def:MDP} using an example. 
\begin{example}\label{ex:road network}
    We consider a road segment consisting of two lanes modeled by a discrete set of locations $\mathcal{L}=\{l_1,\ldots,l_8\}$, as shown in Fig. \ref{fig:example}.
    A set of autonomous vehicles $\mathcal{N}=\{1,2,3,4\}$ shares the road segment. 
    Each state $s_i\in\mathcal{S}_i$ of a vehicle $i$ represents its location, where $\mathcal{S}_i=\mathcal{L}$ for all $i\in\mathcal{N}$.
    Each vehicle $i$ could take certain actions from a set $\mathcal{A}_i$ to transit among the locations.
    Due to the heterogeneous makes, types, and models, the vehicles follow different transition probabilities $Pr_i:\mathcal{S}_i\times\mathcal{A}_i\rightarrow\Delta(\mathcal{S}_i)$, where $Pr(s_i'|s_i,a_i)$ represents the probability of transitioning from a location $s_i$ to its neighboring location $s_i'$ when the vehicle $i$ takes action $a_i$.
    Since the vehicles are transition independent, the joint transition probability of all agents is expressed as $Pr(s'|s,a)=\prod_{i\in\mathcal{N}}Pr_i(s_i'|s_i,a_i)$, where $s'=(s_1,\ldots,s_N)$, $s=(s_1,\ldots,s_N)$, and $a=(a_1,\ldots,a_N)$.
    A path $\rho$ in this example then represents a sequence of joint locations and actions of all vehicles.
    A policy $\pi_i$ of vehicle $i$ specifies the action that should be taken by the vehicle given the current stage $t$ and joint location $s$ of all vehicles.
\end{example}

We assume that there is a subset of unsafe states $\hat{\mathcal{S}}\subset\mathcal{S}$ that should be avoided by the agents.
In addition, we assume that $Pr(s'|s,a)=0$ for all $s'\notin\hat{\mathcal{S}}, s\in\hat{\mathcal{S}}$, and $a\in\mathcal{A}$.
This assumption indicates that once an agent reaches an unsafe state, the safety constraint is violated and cannot be recovered.

In this paper, we focus on the case where an instance of safety violation is observed. 
Without loss of generality, we focus on the case where a finite path $\rho$ of length $T$ is observed such that  $\rho^T\in\hat{\mathcal{S}}$.
We investigate the following problem.

\begin{problem}\label{prob:formulation}
    Consider a collection of agents modeled by an MMDP where a subset of states $\hat{\mathcal{S}}\subset\mathcal{S}$ is unsafe. 
    Suppose a path $\rho$ that reached $\hat{\mathcal{S}}$ is observed.
    Characterize the subset of agents that are responsible for the safety violation. 
    Furthermore, if an agent is considered responsible, quantify to what extent this agent is responsible for the safety violation. 
\end{problem}
We illustrate Problem \ref{prob:formulation} using Example \ref{ex:road network}.
\begin{excont}[Continued]
Suppose the autonomous vehicles shown in Fig. \ref{fig:example} are required to avoid colliding with each other. 
Then, the set of unsafe states $\hat{\mathcal{S}}$ can be expressed as $\hat{\mathcal{S}}=\{(s_1,\ldots,s_N):\exists i,j\in\mathcal{N}\text{ s.t. }s_i=s_j\}$.
Suppose a path $\rho$ is observed, such that there exists a stage $t\in\{0,\ldots,T-1\}$ in which vehicles $2$ (orange) and $3$ (yellow) collide, i.e., $s_2=s_3$ holds at some stage $t$.
Once these vehicles collide, they will stop there and the safety constraint cannot be recovered.
Then the problem of interest is to quantify the vehicles' responsibilities for the collision.
\end{excont}

	\section{Solution Approach}\label{sec:solutionapproach}

This section details development of a concept that we term 
\emph{Degree of Responsibility (DoR)} to measure individual agents' responsibility for the safety violation observed in path $\rho$.

\subsection{Overview}

Our goal is to quantify the responsibility of each individual agent for the safety violation observed in path $\rho$.
We determine responsibility of an agent by examining whether it could have taken an action or actions different than those 
observed in path $\rho$ in order to prevent the safety violation.
Since we only have access to the observed path $\rho$, we lack knowledge of how the agents would behave when they take different actions.
We address this lack of information by constructing a collection of \emph{counterfactual worlds}. Each counterfactual world represents a different hypothetical scenario.

In each counterfactual world, a subset of agents, denoted $\mathcal{Y}\subset\mathcal{N}$, chooses actions from a safe policy, i.e., a policy that minimizes the probability of violating the safety constraint.
We formalize such a counterfactual world by formulating it as a coalitional game \cite{fudenberg1991game}, where a subset of agents forms a coalition $\mathcal{Y}$ and follow the safe policy at a particular stage $t$, while the remaining agents in $\mathcal{N}\setminus\mathcal{Y}$ follow the same actions as they have done in the $t$-th stage of path $\rho$. 
We characterize each counterfactual world by constructing a utility function representing the probability of reaching the unsafe states.

Given the utility functions of the counterfactual worlds, quantifying agents' responsibilities then reduces to a problem of distributing the utilities among the agents.
One solution concept to distribute the utility is via the Shapley value \cite{roth1988shapley}. The Shapley value has been adopted in multiple engineering applications \cite{rozemberczki2022shapley,han2022stable} to evaluate the contribution of individual agents in a coalition.
This motivates us to quantify agent responsibility by splitting utility functions in all counterfactual worlds to agents using the Shapley value.

\subsection{Degree of Responsibility}

In this subsection, we discuss how to construct the collection of counterfactual worlds.
We then propose the concept of DoR to distribute utilities of these counterfactual worlds among agents to measure their responsibilities for safety violation.


We denote the collection of counterfactual worlds as $\{\mathcal{C}_\mathcal{Y}^t:\mathcal{Y}\subseteq\mathcal{N},t\in\{0,\ldots,T-1\}\}$.
In each counterfactual world $\mathcal{C}_\mathcal{Y}^t$, a subset of agents forms a coalition $\mathcal{Y}\subseteq\mathcal{N}$ at stage $t\in\{0,\ldots,T-1\}$ to collaboratively guarantee satisfaction of the safety constraint, i.e., avoid reaching unsafe states $\hat{\mathcal{S}}$.
Each counterfactual world $\mathcal{C}_\mathcal{Y}^t$ is constructed as follows:
\begin{itemize}
    \item The joint state of agents in $\mathcal{C}_\mathcal{Y}^t$, denoted $\tilde{s}^t$, is identical to the $t$-th stage of the observed path $\rho$, i.e., $\tilde{s}^t=\rho^t$.
    \item A subset of agents forms a coalition $\mathcal{Y}$ and takes a certain action $\tilde{a}_i^t=\pi_i(\tilde{s}^t,t)$ at stage $t$ according to a non-stationary policy $\pi_i$ for all $i\in\mathcal{Y}$.
    The policies $\{\pi_i:i\in\mathcal{Y}\}$ minimizes the probability of reaching the unsafe states starting from stage $t$. 
    \item Each agent $j\notin\mathcal{Y}$ follows the same action that it has taken along 
    $\rho$ at stage $t$, i.e., $\tilde{a}_j^t=a_j^t$ for all $j\notin\mathcal{Y}$.
    \item All agents follow a joint non-stationary policy $\tilde{\pi}$ for all stages $t'\in\{t+1,\ldots,T-1\}$ to minimize the probability of reaching the unsafe states.
\end{itemize}

We illustrate the construction of counterfactual worlds by continuing our discussion from Example \ref{ex:road network}.
\begin{excont}[Continued]
    We construct the counterfactual world $\mathcal{C}_{\{1\}}^0$ as follows.
    In this counterfactual world, the joint state of all agents is $\tilde{s}^0=\rho^0$.
    In addition, we have $\mathcal{Y}=\{1\}$.
    Thus, vehicle $1$ will follow policy $\pi_1$ at stage $0$ that might be different from its action taken along the path $\rho$.
    All other vehicles in $\mathcal{N}\setminus\{1\}=\{2,3,4\}$ will follow actions that they have taken at stage $0$ in path $\rho$.
    For stages $t'\in\{1,\ldots,T-1\}$, all agents will follow a joint non-stationary policy that minimizes the probability of reaching the unsafe states.
\end{excont}


There are two major differences between each counterfactual world $\mathcal{C}_\mathcal{Y}^t$ and the observed path $\rho$.
First, each agent $i$ in coalition $\mathcal{Y}$ will take actions following a different policy $\pi_i$.
This allows us to characterize whether safety can be preserved when agents in $\mathcal{Y}$ could have taken some alternative actions.
Moreover, the agents that are not in coalition $\mathcal{Y}$ are allowed to take actions observed at the stage $t$ in the counterfactual world $\mathcal{C}_\mathcal{Y}^t$.
For all stages $t'=t+1,\ldots,T-1$, all agents will follow a policy $\tilde{\pi}$ that minimizes the probability of reaching the unsafe states.
Thus, we can isolate the effect of actions taken by agents in $\mathcal{N}\setminus\mathcal{Y}$ at stage $t$ of the observed path $\rho$.

Leveraging the differences between the counterfactual worlds and the observed path, we develop a utility function for each counterfactual world to characterize the probability of reaching the unsafe states as follows:
\begin{equation}\label{eq:reward}
    r(C_\mathcal{Y}^t;\{\pi_i\}_{i\in\mathcal{Y}},\tilde{\pi})=\min_{\substack{\pi_i:i\in\mathcal{Y}\\\tilde{\pi}}}P(\{t,\ldots,T-1\}\rightsquigarrow\hat{\mathcal{S}}|\mathcal{C}_\mathcal{Y}^t),
\end{equation}
where $\{t,\ldots,T-1\}\rightsquigarrow\hat{\mathcal{S}}$ denotes the event that there exists some stage $t'\in\{t,\ldots,T-1\}$ such that the joint state $\tilde{s}^{t'}\in\hat{\mathcal{S}}$ in the counterfactual world $\mathcal{C}_\mathcal{Y}^t$, and $P(\cdot|\mathcal{C}_\mathcal{Y}^t)$ represents the probability evaluated in the counterfactual world $\mathcal{C}_\mathcal{Y}^t$.
In Eq. \eqref{eq:reward}, the probability of reaching unsafe states is minimized over two policies: (i) the policies $\pi_i$ of each agent $i$ in coalition $\mathcal{Y}$ at stage $t$, and (ii) the policies $\Tilde{\pi}$ of all agents $i\in\mathcal{N}$ at stages $t'\in\{t+1,\ldots,T-1\}$.
Therefore, the utility function in Eq. \eqref{eq:reward} characterizes the probability of reaching unsafe states if agents in a coalition $\mathcal{Y}$ could have taken some alternative actions. 
Consequently, this allows us to quantify to what extent actions taken by agents outside the coalition, i.e., in $\mathcal{N}\setminus\mathcal{Y}$ at stage $t$ of path $\rho$ contributes to the observed safety violation.

We characterize the utility function in Eq. \eqref{eq:reward} as follows.
\begin{proposition}\label{prop:monotone}
    Let $\mathcal{C}_\mathcal{Y}^t$ and $\mathcal{C}_{\mathcal{Y}'}^t$ be two counterfactual worlds such that $\mathcal{Y}\subseteq \mathcal{Y}'$.
    Then 
    \begin{equation}\label{eq:monotone}
        r(C_\mathcal{Y}^t;\{\pi_i\}_{i\in\mathcal{Y}},\tilde{\pi})\geq r(C_{\mathcal{Y}'}^t;\{\pi_i'\}_{i\in\mathcal{Y}},\tilde{\pi}'),
    \end{equation}
    where $\{\pi_i\}_{i\in\mathcal{Y}}$ and $\Tilde{\pi}$ are associated with $\mathcal{C}_\mathcal{Y}^t$, and $\{\pi_i'\}_{i\in\mathcal{Y}}$ and $\Tilde{\pi}'$ are associated with $\mathcal{C}_{\mathcal{Y}'}^t$.
\end{proposition}

We note that the counterfactual worlds are independent of each other.
Therefore, given the utility function in Eq. \eqref{eq:reward}, we can characterize the probability of reaching the unsafe states $\hat{\mathcal{S}}$ 
if some agents in $\mathcal{Y}$ could have taken some alternative actions at all stages as follows:
\begin{equation}\label{eq:utility}
    u(\mathcal{Y}) = \sum_{t=0}^{T-1}r(C_\mathcal{Y}^t;\{\pi_i\}_{i\in\mathcal{Y}},\tilde{\pi}),
\end{equation}
Eq. \eqref{eq:utility} can then be utilized to quantify contribution of actions taken by agents in $\mathcal{N}\setminus\mathcal{Y}$ at all stages $t=0,\ldots,T-1$ in path $\rho$ towards the observed safety violation.

Given Eq. \eqref{eq:utility}, we use the Shapley value \cite{roth1988shapley} to distribute the total amount of utility $u(\mathcal{Y})$ among agents as follows:
\begin{equation}\label{eq:phi}
    \phi_i = \frac{1}{|\mathcal{N}|}\sum_{\mathcal{Y}\subseteq\mathcal{N}\setminus\{i\}}{\mathcal{N}\choose |\mathcal{N}|-|\mathcal{Y}|-1}^{-1}(u(\mathcal{Y}\cup\{i\})-u(\mathcal{Y})).
\end{equation}
Eq. \eqref{eq:phi} considers all possible orders in which the agents can join a coalition, and distributes the utility to agent $i$ as its marginal contribution, i.e., $u(\mathcal{Y}\cup\{i\})-u(\mathcal{Y})$, averaged over all possible orders.

We finally normalize $\phi_i$ among all the agents, and define the normalized value as the \emph{\textbf{degree of responsibility (DoR)}} of agent $i$, which is given as
\begin{equation}\label{eq:dor}
    \psi_i = \frac{\phi_i}{\sum_{i\in\mathcal{N}}\phi_i}.
\end{equation}
The DoR can then be used to quantify the responsibility of agent $i$ for violating the safety constraint.
By Proposition \ref{prop:monotone}, we have that $\phi_i$ in Eq. \eqref{eq:phi} is non-positive for all agent $i\in\mathcal{N}$. Hence the DoRs satisfy that $\psi_i\in[0,1]$ for all agent $i\in\mathcal{N}$ and $\sum_{i\in\mathcal{N}}\psi_i=1$.
This allows us to use the DoR to compare and rank the responsibilities of different agents for the safety violation. 
We 
illustrate the 
DoR using Example \ref{ex:road network}.
\begin{excont}[Continued]
    We consider the set of counterfactual worlds $\{\mathcal{C}_{\{1\}}^t\}_{t=0}^{T-1}$. 
    When $\mathcal{Y}=\{1\}$, Eq. \eqref{eq:utility} models the probability of reaching unsafe states in these counterfactual worlds.
    Note that the vehicle $1$ follows the policy $\pi_1(\rho^t,t)$ at stage $t$ in each counterfactual world $\mathcal{C}_{\{1\}}^t$.
    Therefore, Eq. \eqref{eq:utility} characterizes the influence of actions taken by vehicle $1$ at different stages $t=0,\ldots,T-1$. 
    In Eq. \eqref{eq:phi}, the term $u(\mathcal{Y}\cup\{2\})-u(\mathcal{Y})$ quantifies the effect of actions taken by vehicle $2$ at different stages, given that vehicle $1$ will follow policy $\pi_1$ at each stage.
    By taking all possible permutations $\mathcal{Y}\subseteq\mathcal{N}\setminus\{2\}$ into consideration and normalizing among all agents, Eq. \eqref{eq:dor} yields the DoR of vehicle $2$ to be $0.5$. 
\end{excont}

\subsection{Computation of Policies in the Counterfactual World}

To compute the DoRs for all agents, Eq. \eqref{eq:reward}-\eqref{eq:dor} require to (i) construct a policy $\pi_i$ for each agent $i\in\mathcal{Y}$ at stage $t$ that minimizes the probability of reaching unsafe states $\hat{\mathcal{S}}$, (ii) construct a policy $\Tilde{\pi}$ for all agents at stages $t'=t+1,\ldots,T-1$, and (ii) evaluate the probability of reaching the unsafe states $\hat{\mathcal{S}}$ in each counterfactual world.
We refer to the policies $\pi_i$ followed by the agents in $\mathcal{Y}$ and $\Tilde{\pi}$ in the counterfactual worlds as the \emph{safe policies}.
In what follows, we discuss how to compute the safe policies and how to evaluate the probability of reaching the unsafe states in each counterfactual world.


To compute safe policies and evaluate the probability of reaching the unsafe states $\hat{\mathcal{S}}$ in each counterfactual world, we define a Q-function for any given joint state-action pair as:
\begin{align}
    Q(s,a,0) &= \begin{cases}
        0, &\forall a \mbox{ if } s\notin\hat{\mathcal{S}}\\
        1, &\forall a \mbox{ if } s\in\hat{\mathcal{S}}
    \end{cases},\label{eq:Q-value 1}\\
    Q(s,a,t) &= \sum_{s'\in\mathcal{S}}Pr(s'|s,a)\min_{a'}Q(s',a',t-1),~\forall t>1.\label{eq:Q-value 2}
\end{align}
The Q-function specifies the probability of reaching the unsafe states in $t$ stages when the agents start from state $s$ and take joint action $a$.
By the definition of the Q-function, we have that the policy $\pi_i$ for each agent $i\in\mathcal{Y}$ can be computed by minimizing $Q(\rho^t,\Tilde{a}^t,T-t)$ over $\Tilde{a}^t$, where the $q$-th entry of $\Tilde{a}^t$ is $\pi_q(\rho^t,t)$ if $q\in\mathcal{Y}$, and $a^t$ otherwise.
The safe policy $\tilde{\pi}$ for stage $t'\in\{t+1,\ldots,T-1\}$ can be computed as $\tilde{\pi}=\argmin_{\Tilde{\pi}'}Q(s,\Tilde{\pi}',T-t)$ for any joint state $s$.

Using the definitions of Q-function and safe policies, the probability of reaching the unsafe states in the counterfactual world $\mathcal{C}_\mathcal{Y}^t$ can be represented as
\begin{equation}\label{eq:Q replacement}
    r(C_\mathcal{Y}^t;\{\pi_i\}_{i\in\mathcal{Y}},\tilde{\pi})
    = Q(\rho^t, \tilde{a}^t, T-t),
\end{equation}
where $\Tilde{a}^t = (\Tilde{a}_1^t,\ldots,\tilde{a}_N^t)$ is the joint action taken by all agents in counterfactual world $\mathcal{C}_\mathcal{Y}^t$ when joint state is $\rho^t$ at stage $t$.

    \section{Efficient Computation of DoR}\label{sec:efficientcomputation}

As the number of agents in the MMDP increases, the computational complexity of determining DoRs grows exponentially, rendering it intractable for large-scale multi-agent CPS.
In particular, we identify two key factors contributing to the high computational complexity of calculating DoR. 

The first factor that increases the computational complexity of calculating DoR is that Eq. \eqref{eq:phi} requires evaluating all possible permutations of agents in $\mathcal{Y}$.
Furthermore, for each permutation, the value of $\phi_i$ needs to be computed for each individual agent $i$.
The second factor is that the joint state and action spaces of the MMDP grow exponentially with the number of agents. 
Consequently, there are a total of $T\prod_{i=1}^N|\mathcal{S}_i|\prod_{i=1}^N|\mathcal{A}_i|$ number of Q-functions that need to be calculated in Eq. \eqref{eq:Q-value 1}-\eqref{eq:Q-value 2} in order to apply 
Eq. \eqref{eq:Q replacement}.

In what follows, we first develop a heuristic algorithm to efficiently compute DoRs for arbitrary MMDPs.
We then demonstrate that DoRs can be computed efficiently when the MMDP $\mathbb{M}$ possesses special structural properties. 

{We defer the proofs of our results to an extended version due to space constraints.}

\subsection{Heuristic Algorithm for Efficient Computation of DoR}

In this subsection, we mitigate the computational complexity of our approach by developing an algorithm that identifies a subset of agents that are likely to be responsible for the observed path $\rho$ reaching the unsafe states $\hat{\mathcal{S}}$.
We denote the subset of responsible agents as $\mathcal{R}\subseteq\mathcal{N}$.
By identifying the set $\mathcal{R}$, the value of $\phi_i$ can then be calculated as 
\begin{equation}\label{eq:improved phi}
    \phi_i = \frac{1}{|\mathcal{N}|}\sum_{\mathcal{Y}\subseteq\mathcal{R}\setminus\{i\}}{\mathcal{R}\choose |\mathcal{R}|-|\mathcal{Y}|-1}^{-1}(u(\mathcal{Y}\cup\{i\})-u(\mathcal{Y})),
\end{equation}
which significantly reduces the number of permutations that need to be evaluated, i.e., $2^{|\mathcal{R}|-1}$ coalitions in Eq. \eqref{eq:improved phi} compared with $2^{|\mathcal{N}|-1}$ coalitions in Eq. \eqref{eq:phi}.
In the worst-case where $\mathcal{R}=\mathcal{N}$, the computational complexity of evaluating $\phi_i$ using Eq. \eqref{eq:improved phi} is the same as using Eq. \eqref{eq:phi}.

The key insight enabling identification of responsible agents is development of a concept named $\epsilon$-marginally safe policy. 
\begin{definition}[$\epsilon$-Marginally Safe Policy]\label{def:marginal safe policy}
    Given a joint state $s^t$ and joint action $a^t=(a_1^t,\ldots,a_N^t)$ at stage $t$, a policy $\pi_i$ is said to be an $\epsilon$-marginally safe policy for agent $i$ if 
    \begin{equation}\label{eq:marginal safe policy}
        Q(s^t,a^t,T-t)-Q(s^t,\hat{a}^t,T-t)\geq \epsilon,
    \end{equation}
     where  $\hat{a}^t=(a_1^t,\ldots,a_{i-1}^t,\pi_i(s^t),a_{i+1}^t,\ldots,a_N^t)$.
\end{definition}
We use Definition \ref{def:marginal safe policy} and identify the set of responsible agents in Algorithm \ref{algo:identify}.
The algorithm takes MMDP $\mathbb{M}$, observed path $\rho=s^0, a^0, s^1,\ldots, a^{T-2}, s^{T-1}$, and a threshold $\epsilon$ as inputs.
It first initializes the set of responsible agents as $\mathcal{R}=\emptyset$, and then iterates over all agents $i\in\mathcal{N}$. 
At each iteration, the algorithm computes an $\epsilon$-marginally safe policy for each agent $i$, assuming that other agents follow the same actions $a_j^t$ at state $\rho^t$ for all $j\in\mathcal{N}\setminus\{i\}$.
Existence of an $\epsilon$-marginally safe policy for an agent $i$ at stage $t$ can be verified by searching for an action $\hat{a}_i^t$ such that Eq. \eqref{eq:marginal safe policy} holds.
In the worst-case, computational complexity of searching for an $\epsilon$-marginally safe policy of agent $i$ is $\mathcal{O}(|\mathcal{A}_i|)$ at each stage $t$.
Therefore, the computational complexity of Algorithm \ref{algo:identify} is $\max_{i\in\mathcal{N}}TN\mathcal{O}(|\mathcal{A}_i|)$, which is linear in number of agents. 

\begin{center}
  	\begin{algorithm}[!htp]
  		\caption{Identification of Responsible Agents $\mathcal{R}$}
  		\label{algo:identify}
  		\begin{algorithmic}[1]
  			\State \textbf{Input}: MMDP $\mathbb{M}$, observed path $\rho$, threshold $\epsilon$ 
  			\State \textbf{Output:} The set of responsible agents $\mathcal{R}\subseteq\mathcal{N}$
            \State \textbf{Initialize:} $\mathcal{R}\leftarrow\emptyset$
            
            \For{$i\in\mathcal{N}$}
            \For{$t\in\{0,\ldots,T-1\}$}
            \State Let each agent $j\in\mathcal{N}\setminus\{i\}$ take the action $a_j^t$
            \State Search for an action $\hat{a}_i^t\neq a_i^t$ for agent $i$ such that Eq. \eqref{eq:marginal safe policy} holds
            \If{ there exists such an action $\hat{a}_i^t$} 
            \State $\mathcal{R}\leftarrow\mathcal{R}\cup\{i\}$
            \EndIf
            \EndFor
            \EndFor
            \State \textbf{Return} $\mathcal{R}$
  		\end{algorithmic}
  	\end{algorithm}
\end{center}

\subsection{Structural MMDP Property for Efficient DoR Computation}

We demonstrate that when the MMDP $\mathbb{M}$ satisfies certain structural properties, the DoRs can be calculated efficiently.

In the remainder of this section, we assume that interactions among agents are governed by an underlying undirected graph $\mathcal{G}=(\mathcal{N},\mathcal{E})$, where $\mathcal{N}$ is the set of nodes representing agents, and $\mathcal{E}\subset\mathcal{N}\times\mathcal{N}$ is a set of edges.
We denote the set of neighboring agents of an agent $i$ as $\mathcal{B}_i$, i.e., for any agent $j\in\mathcal{B}_i$, there exists an edge $e_{ij}\in\mathcal{E}$ connecting agents $i$ and $j$.
The $k$-hop neighbors of an agent $i$ is denoted as $\mathcal{B}_i^k$.
We also define the states of the agents within and without the $k$-hop neighbors of agent $i$ as $s_i^{\mathcal{B}^k}$ and $s_{-i}^{\mathcal{B}^k}$, respectively.
Actions of the agents within and without the $k$-hop neighbors of agent $i$ are denoted similarly as $a_i^{\mathcal{B}^k}$ and $a_{-i}^{\mathcal{B}^k}$, respectively.
Therefore, for any joint state $s\in\mathcal{S}$ and joint action $a\in\mathcal{A}$, they can be rewritten as $s=(s_i^{\mathcal{B}^k},s_{-i}^{\mathcal{B}^k})$ and $a=(a_i^{\mathcal{B}^k},a_{-i}^{\mathcal{B}^k})$.
For any joint state $s\in\mathcal{S}$ and action $a\in\mathcal{A}$, we assume that the transition probability $Pr$ in Definition \ref{def:MDP} can be represented as \cite{qu2020scalable}
\begin{equation}\label{eq:local dependence}
    Pr(s'|s,a) = \prod_{i\in\mathcal{N}}Pr_i(s_i'|s_i^\mathcal{B},a_i),
\end{equation}
where $s_i^\mathcal{B}$ represents the joint states of agent $i$ and its neighboring agents, and $Pr_i$ is the transition function of agent $i$.

In what follows, we assume that the MMDP $\mathbb{M}$ satisfies an exponential decay property, which is defined below.

\begin{definition}[$(c,\gamma)$ Exponential Decay Property \cite{qu2020scalable}]\label{def:exponential decay relaxed}
    The $(c,\gamma)$ exponential decay property holds for some $c>0$ and $\gamma\in(0,1)$ if the following inequality holds
    \begin{multline}
        \Big|Q\left((s_i^{\mathcal{B}^k},s_{-i}^{\mathcal{B}^k}),(a_i^{\mathcal{B}^k},a_{-i}^{\mathcal{B}^k}),0\right) \\
        - Q\left((s_i^{\mathcal{B}^k},s_{-i}^{\mathcal{B}^{k'}}),(a_i^{\mathcal{B}^k},a_{-i}^{\mathcal{B}^{k'}}),0\right)\Big|
        \leq c\gamma^{k+1}
    \end{multline}
    for any agent $i\in\mathcal{N}$, joint states $(s_i^{\mathcal{B}^k},s_{-i}^{\mathcal{B}^k}), (s_i^{\mathcal{B}^k},s_{-i}^{\mathcal{B}^{k'}})\in\mathcal{S}$, and joint actions $(a_i^{\mathcal{B}^k},a_{-i}^{\mathcal{B}^k}), (a_i^{\mathcal{B}^k},a_{-i}^{\mathcal{B}^{k'}})\in\mathcal{A}$.
\end{definition}
MMDPs that satisfy the $(c, \gamma)$ exponential decay property 
are commonly encountered in practical scenarios. 
For example, when a set of 
vehicles collides with an obstacle, it is improbable that a vehicle located far away from the collision site will be responsible for the safety violation.
This is because impact of maneuvers executed by the distant vehicle diminishes rapidly as it is spatially separated from the 
collision, consistent with the $(c, \gamma)$ exponential decay property.

Given Definition \ref{def:exponential decay relaxed}, we can approximate the Q-function using a local Q-function, denoted as $Q^L$, as follows:
\begin{multline}\label{eq:local Q}
    Q^L(s_i^{\mathcal{B}^k},a_i^{\mathcal{B}^k},0) \\
    = \sum_{s_{-i}^{\mathcal{B}^k}\in\mathcal{S}_{-i}^{\mathcal{B}^k},a_{-i}^{\mathcal{B}^k}\in\mathcal{A}_{-i}^{\mathcal{B}^k}}\omega_i\left((s_i^{\mathcal{B}^k},s_{-i}^{\mathcal{B}^k}),(a_i^{\mathcal{B}^k},a_{-i}^{\mathcal{B}^k})\right)\\
    \cdot Q\left((s_i^{\mathcal{B}^k},s_{-i}^{\mathcal{B}^k}),(a_i^{\mathcal{B}^k},a_{-i}^{\mathcal{B}^k}),0\right),
\end{multline}
where $\omega_i\left((s_i^{\mathcal{B}^k},s_{-i}^{\mathcal{B}^k}),(a_i^{\mathcal{B}^k},a_{-i}^{\mathcal{B}^k})\right)$ are non-negative weights such that
\begin{equation}\label{eq:weight}
    \sum_{s_{-i}^{\mathcal{B}^k}\in\mathcal{S}_{-i}^{\mathcal{B}^k},a_{-i}^{\mathcal{B}^k}\in\mathcal{A}_{-i}^{\mathcal{B}^k}}\omega_i\left((s_i^{\mathcal{B}^k},s_{-i}^{\mathcal{B}^k}),(a_i^{\mathcal{B}^k},a_{-i}^{\mathcal{B}^k})\right)=1,
\end{equation}
$\mathcal{S}_{-i}^{\mathcal{B}^k}$ is the set of states of all agents outside the $k$-hop neighborhood of agent $i$, and $\mathcal{A}_{-i}^{\mathcal{B}^k}$ is the set of actions of all agents outside the $k$-hop neighborhood of agent $i$.
We have the following optimality guarantee when approximating the Q-function using the local Q-function in Eq. \eqref{eq:local Q}.
\begin{lemma}\label{lemma:local Q}
    Let $s=(s_i^{\mathcal{B}^k},s_{-i}^{\mathcal{B}^{k'}})\in\mathcal{S}$ and $a=(a_i^{\mathcal{B}^k},a_{-i}^{\mathcal{B}^{k'}})\in\mathcal{A}$ be the joint state and action of the MMDP $\mathbb{M}$. 
    If the MMDP $\mathbb{M}$ satisfies the $(c,\gamma)$ exponential decay property in Definition \ref{def:exponential decay relaxed}, then the following relationship holds
    \begin{equation*}
        \sup_{s_{-i}^{\mathcal{B}^{k'}}\in\mathcal{S}_{-i}^{\mathcal{B}^{k'}},a_{-i}^{\mathcal{B}^{k'}}\in\mathcal{A}_{-i}^{\mathcal{B}^k}}|Q^L(s_i^{\mathcal{B}^k},a_i^{\mathcal{B}^k},0)-Q(s,a,0)|\leq c\gamma^{k+1}.
    \end{equation*}
\end{lemma}

Lemma \ref{lemma:local Q} shows that we can approximate the $0$-stage-to-go Q-function using only the local Q-function, which requires tracking joint states and actions of a subset of agents, i.e., agents within the $k$-hop neighborhood of agent $i$.
By Lemma \ref{lemma:local Q}, we can approximate the Q-function using local Q-functions for arbitrary 
$t\in\{1,\ldots,T-1\}$, 
and thus approximate safe policies for each agent in a distributed manner as follows:

\begin{multline}\label{eq:local Q update}
    Q^L(s_i^{\mathcal{B}^k},a_i^{\mathcal{B}^k},t) = \sum_{s_i^{\mathcal{B}^{k'}}\in \mathcal{S}_i^{\mathcal{B}^{k}}}Pr^L(s_i^{\mathcal{B}^{k'}}|s_i^{\mathcal{B}^k},a_i^{\mathcal{B}^k})\\
    \cdot\min_{a_i^{\mathcal{B}^{k'}}}Q^L(s_i^{\mathcal{B}^k},a_i^{\mathcal{B}^{k'}},t-1),~\forall t\in\{1,\ldots,T-1\},
\end{multline}
where $Pr^L(s_i^{\mathcal{B}^{k'}}|s_i^{\mathcal{B}^k},a_i^{\mathcal{B}^k})$ is the marginalized transition probability given as 
\begin{multline}\label{eq:marginal transition prob}
Pr^L\left(s_i^{\mathcal{B}^{k'}}|s_i^{\mathcal{B}^k},a_i^{\mathcal{B}^k}\right) = \\
    \sum_{\substack{s_{-i}^{\mathcal{B}^k},s_{-i}^{\mathcal{B}^{k'}}\in\mathcal{S}_{-i}^{\mathcal{B}^k} \\
    a_{-i}^{\mathcal{B}^k}\in\mathcal{A}_{-i}^{\mathcal{B}^k}}}Pr\left((s_i^{\mathcal{B}^{k'}},s_{-i}^{\mathcal{B}^{k'}})|(s_i^{\mathcal{B}^k},s_{-i}^{\mathcal{B}^{k}}), (a_i^{\mathcal{B}^k},a_{-i}^{\mathcal{B}^{k}})\right).
\end{multline}

The optimality guarantee of the local Q-function obtained in Eq. \eqref{eq:local Q update} is given in the following result.
\begin{proposition}\label{prop:local Q}
    Let $s=(s_i^{\mathcal{B}^k},s_{-i}^{\mathcal{B}^{k'}})\in\mathcal{S}$ and $a=(a_i^{\mathcal{B}^k},a_{-i}^{\mathcal{B}^{k'}})\in\mathcal{A}$ be the joint state and action of the MMDP $\mathbb{M}$. 
    If the MMDP $\mathbb{M}$ satisfies the $(c,\gamma)$ exponential decay property in Definition \ref{def:exponential decay relaxed}, then for any stage $t=0,\ldots,T-1$, the following relationship holds
    \begin{equation*}
        \sup_{s_{-i}^{\mathcal{B}^{k'}}\in\mathcal{S}_{-i}^{\mathcal{B}^{k'}},a_{-i}^{\mathcal{B}^{k'}}\in\mathcal{A}_{-i}^{\mathcal{B}^k}}|Q^L(s_i^{\mathcal{B}^k},a_i^{\mathcal{B}^k},t)-Q(s,a,t)|\leq c\gamma^{k+1}.
    \end{equation*}
\end{proposition}
We note that we only need to evaluate the local Q-function on a smaller set of joint states and actions.
This allows us to utilize much less memory to store the local Q-function and less computational power to calculate the approximately safe policies, making our solution technique more tractable for large-scale multi-agent CPS.
We remark that the optimality bound in Lemma \ref{lemma:local Q} decreases exponentially as the state and action of more agents are revealed.
We finally present the optimality bound when approximating $\phi_i$ in Eq. \eqref{eq:phi} using the local Q-function as follows.
\begin{theorem}\label{thm:DoR bound}
    Assume that the MMDP $\mathbb{M}$ satisfies the $(c,\gamma)$ exponential decay property in Definition \ref{def:exponential decay relaxed}.
    For an agent $i$, let $\phi_i$ be defined in Eq. \eqref{eq:phi} and evaluated using Eq. \eqref{eq:Q replacement}.
    Let the probability of reaching the unsafe states in each counterfactual world $\mathcal{C}_\mathcal{Y}^t$ be approximated as 
    \begin{equation*}
    r(C_\mathcal{Y}^t;\{\pi_i\}_{i\in\mathcal{Y}},\tilde{\pi})
    \approx Q^L(s_i^{\mathcal{B}^k},a_i^{\mathcal{B}^k},T-t),
    \end{equation*}
    and $\phi_i^L$ be the associated utility distributed to agent $i$ within the coalition $\mathcal{Y}$.
    We have $|\phi_i-\phi_i^L|\leq c^L\gamma^{k+1}$ for all agent $i\in\mathcal{N}$, where
    \begin{equation*}
        c^L = 2\frac{1}{|\mathcal{N}|}\sum_{\mathcal{Y}\subseteq\mathcal{N}\setminus\{i\}}{\mathcal{N}\choose |\mathcal{N}|-|\mathcal{Y}|-1}^{-1}c.
    \end{equation*}
\end{theorem}
Theorem \ref{thm:DoR bound} shows that if the MMDP $\mathbb{M}$ satisfies the $(c,\gamma)$ exponential decay property, then we can use the local Q-function to approximate the safe policies and calculate the DoR for each agent.
The approximation incurs a bounded error as shown in Theorem \ref{thm:DoR bound}.
Furthermore, the error decreases exponentially when the local Q-function utilizes the states from more agents to approximate the Q-function.

\begin{figure*}
\centering
\begin{subfigure}{.3\textwidth}
    \centering
    \includegraphics[width=.65\linewidth]{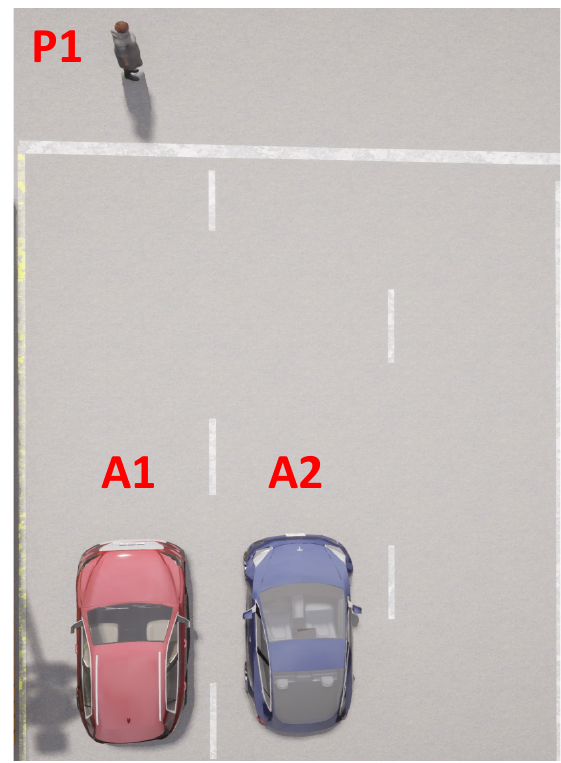}  
    \caption{Initial locations of vehicles}
    \label{fig:s1init_state}
\end{subfigure}
\begin{subfigure}{.3\textwidth}
    \centering
    \includegraphics[width=.65\linewidth]{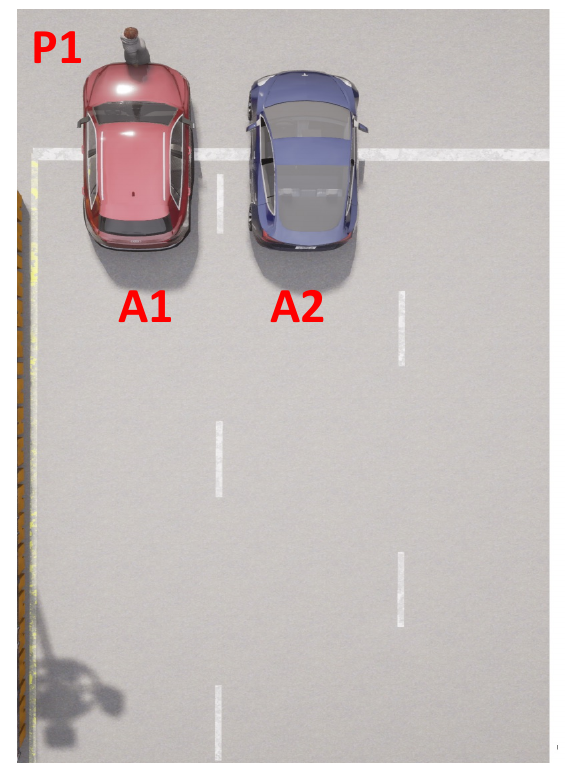}  
    \caption{Illustration of the collision}
    \label{fig:s1collision_state}
\end{subfigure}
\begin{subfigure}{.3\textwidth}
    \centering
    \includegraphics[width=.63\linewidth]{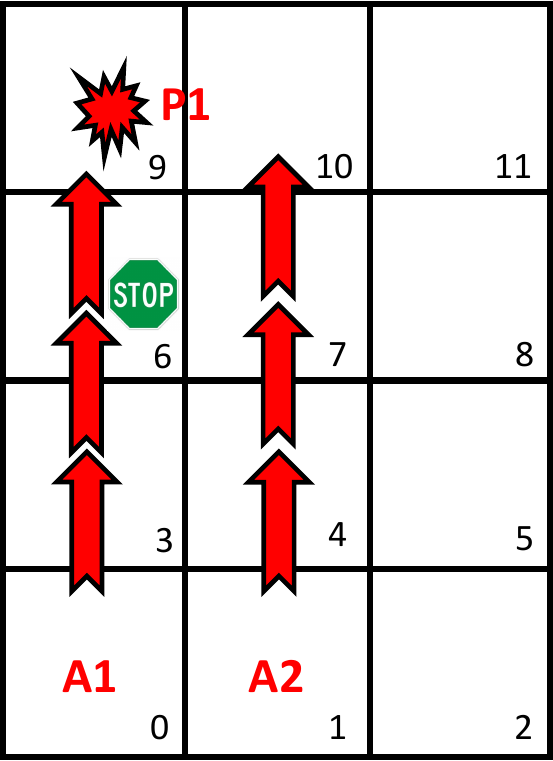}  
    \caption{Illustration of the discretized locations}
    \label{fig:s1grid}
\end{subfigure}
\caption{\textbf{Scenario 1}: This figure presents the initial locations of vehicles, the collision, and the discretized locations for Scenario 1.
The red car is Agent 1 and the blue car is Agent 2.
The initial locations of Agents 1 and 2 (labeled as A1 and A2) are presented in Fig. \ref{fig:s1}-(a).
Fig. \ref{fig:s1}-(b) illustrates a collision, where Agent 1 collides with a pedestrian (P1).
The road segment is discretized into 12 locations $\mathcal{L}=\{0,\ldots,11\}$ as shown in Fig. \ref{fig:s1}-(c). Red arrows represent paths observed in path $\rho$, and the green STOP sign represents the safe policy of Agent 1. Here the STOP sign indicates that the agent should take the stop action to avoid safety violation.
}
\vspace{-1.5em}
\label{fig:s1}
\end{figure*}

\begin{figure*}
\centering
\begin{subfigure}{.3\textwidth}
    \centering
    \includegraphics[width=.95\linewidth]{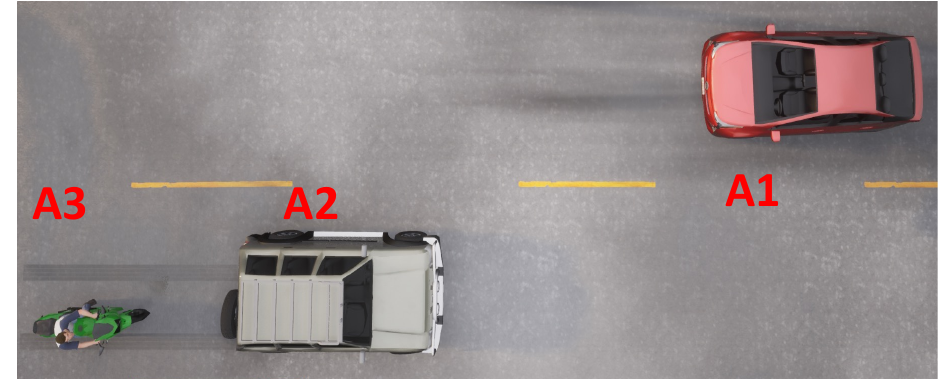}  
    \caption{Initial locations of vehicles}
    \label{fig:s2init_state}
\end{subfigure}
\begin{subfigure}{.3\textwidth}
    \centering
    \includegraphics[width=.95\linewidth]{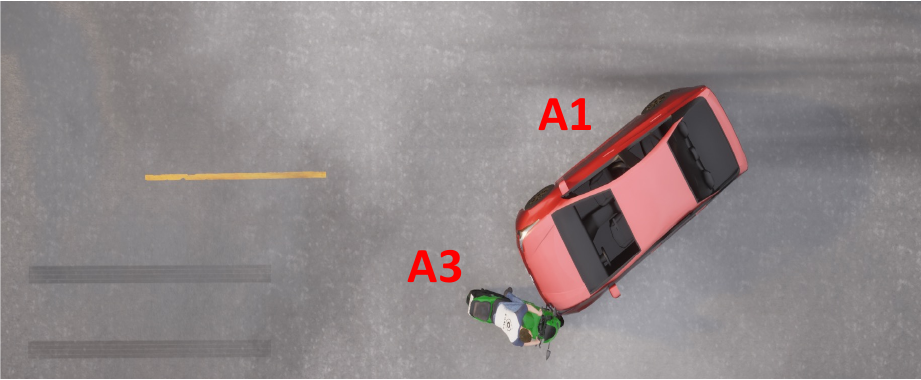}  
    \caption{Illustration of the collision}
    \label{fig:s2collision_state}
\end{subfigure}
\begin{subfigure}{.3\textwidth}
    \centering
    \includegraphics[width=.93\linewidth]{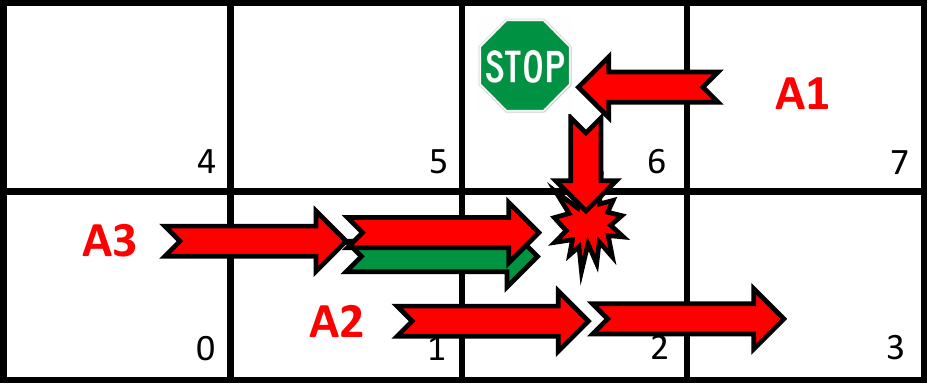}  
    \caption{Illustration of the discretized locations}
    \label{fig:s2grid}
\end{subfigure}
\caption{\textbf{Scenario 2}: This figure presents the initial locations of vehicles, the collision, and the discretized locations for Scenario 2.
The red car is Agent 1, the white SUV is Agent 2, and the green motorcycle is Agent 3.
The initial locations of Agents 1, 2, and 3 (labeled as A1, A2, and A3) are presented in Fig. \ref{fig:s2}-(a).
Fig. \ref{fig:s2}-(b) illustrates the collision, where Agent 1 collides with Agent 3.
The road segment is discretized into 8 locations $\mathcal{L}=\{0,\ldots,7\}$, as shown in Fig. \ref{fig:s2}-(c). Red arrows represent the paths observed in path $\rho$, and the green STOP sign represents the safe policy of Agents 1 and 3.
Here the STOP sign indicates that the agent should take the stop action to avoid safety violation.}
\vspace{-1.5em}
\label{fig:s2}
\end{figure*}

\begin{figure*}
\centering
\begin{subfigure}{.3\textwidth}
    \centering
    \includegraphics[width=.75\linewidth]{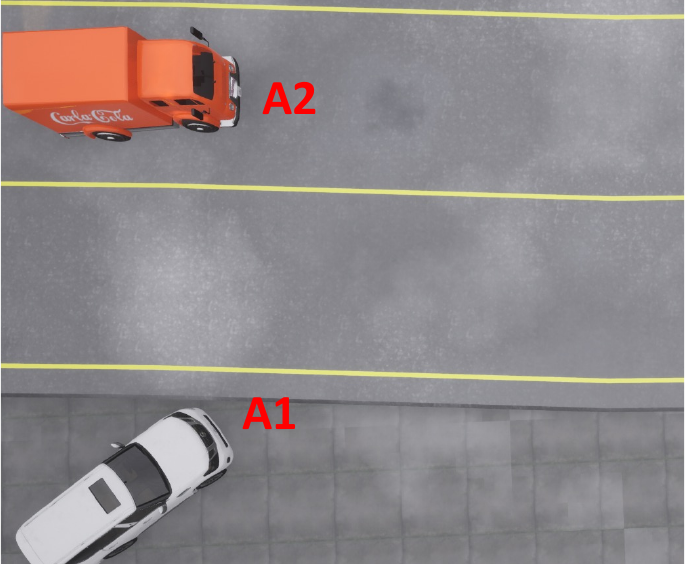}  
    \caption{Initial locations of vehicles}
    \label{fig:s3init_state}
\end{subfigure}
\begin{subfigure}{.3\textwidth}
    \centering
    \includegraphics[width=.75\linewidth]{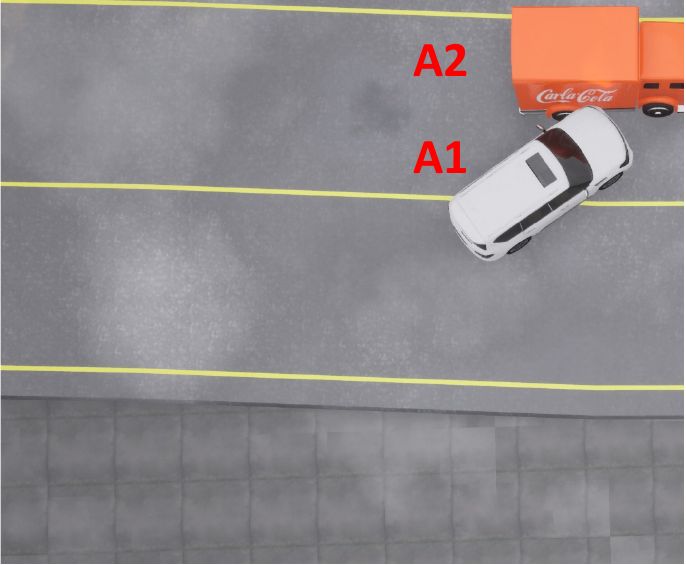}  
    \caption{Illustration of the collision}
    \label{fig:s3collision_state}
\end{subfigure}
\begin{subfigure}{.3\textwidth}
    \centering
    \includegraphics[width=.75\linewidth]{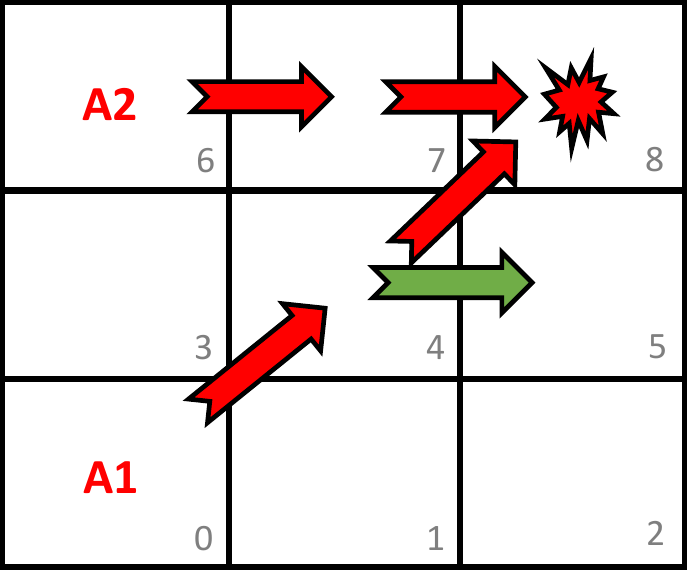}  
    \caption{Illustration of the discretized locations}
    \label{fig:s3grid}
\end{subfigure}
\caption{\textbf{Scenario 3}: This figure presents the initial locations of vehicles, the collision, and the discretized locations for Scenario 3.
The white SUV is Agent 1, and the orange truck is Agent 2.
The initial locations of Agents 1 and 2 (labeled as A1 and A2) are presented in Fig. \ref{fig:s3}-(a).
Fig. \ref{fig:s3}-(b) illustrates the collision, where Agent 1 collides with Agent 2.
The road segment is discretized into 9 locations $\mathcal{L}=\{0,\ldots,8\}$ as shown in Fig. \ref{fig:s3}-(c). Red arrows represent the paths observed in path $\rho$, and the green arrow represents the safe policy of Agent 1.}
\vspace{-1em}
\label{fig:s3}
\end{figure*}

\section{Experiments}\label{sec:experiments}

In this section, we consider three instances of safety violations (i.e., an accident or crash) available from the National Highway Traffic Safety Administration (NHTSA) \cite{crashscenario}. 
Each instance includes information about vehicle movements and dynamics, as well as critical events immediately preceding a crash. 
We render these scenarios using CARLA, an open-source urban driving simulator \cite{dosovitskiy2017carla}. 


We conducted our experiments using a workstation installed with the Linux 5.19.0-43-generic operating system. The workstation is equipped with an Intel\textsuperscript{\tiny\textregistered} Xeon\textsuperscript{\tiny\textregistered} W-2145 CPU @ 3.70 GHz processor, two 16 GB NVIDIA GeForce RTX 2080 Ti GPUs, and 128 GB of memory. We used cvxpy version 0.9.14 within a Python 3.7 environment to construct three  road safety violation scenarios.  
We used the Python MDP Toolbox \cite{heins2022pymdp} to determine the safe policies for vehicle agents after modeling each scenario as an MMDP in Definition \ref{def:MDP}.


{
We first provide details about representation of the three scenarios using CARLA and describe how to construct the MMDP using data collected from CARLA for each scenarios.
Next, we present our experimental results and discussions related to these simulated scenarios. 
Finally, we discuss some scenarios that are not addressed by our present DoR formulation, but nevertheless form a basis for future research.   
}

\subsection{CARLA Simulated Scenarios}

We demonstrate our approach using three road safety violation scenarios derived from \cite{crashscenario}, and simulated in CARLA.

\noindent{\bf Scenario 1:} Our first scenario consists of two vehicles and one non-moving pedestrian in a three-lane road segment. Vehicle Agents 1 and 2 drive parallel to each other and Agent 1 collides with the pedestrian. Snapshots of the initial locations and collision are illustrated in Fig.~\ref{fig:s1}-(a)~and-(b), respectively. 

\noindent{\bf Scenario 2:}
Our second scenario consists of three vehicles in a two-lane road segment.
Vehicle Agents 2 and 3 operate in the same lane, and move in the same direction.
Agent 1, which is operating in the opposite lane to Agents 2 and 3, takes a U-turn.
When Agent 1 is completing the U-turn, it collides with Agent 3.
Snapshots of the initial locations of all vehicles and the collision are presented in Fig.~\ref{fig:s2}-(a)~and-(b), respectively. 

\noindent{\bf Scenario 3:} 
Our third scenario has two vehicles. 
Vehicle Agent 2 stays within its current lane on the highway.
Vehicle Agent 1 departs from the gas station and aims to merge into the lane in which Agent 2 is present.
Agents 1 and 2 collide with each other when Agent 1 tries to merge. 
Snapshots of initial locations and the collision are illustrated in Fig.~\ref{fig:s3}-(a)~and-(b). 

\noindent {\bf MMDP Construction:} 
We discuss how to construct an MMDP for each scenario described above.
We first collect a video stream of length $10$ seconds prior to the traffic accident from the log file available from the CARLA simulator.
This video can then be used to obtain time-stamped trajectories of vehicles and status of all traffic lights.
In real-world applications, video stream captured by RGB-D cameras can be used to estimate vehicle trajectories and signal status \cite{shan2017vehicle}.

We then discretize the environment containing all trajectories into a set of discrete locations $\mathcal{L}$, where each location is represented by a cell.
The length and width of each cell are chosen as the length of the vehicles and width of the lanes, respectively. 
Discretized locations for Scenarios 1, 2, and 3 are illustrated in Fig. \ref{fig:s1}-(c), Fig. \ref{fig:s2}-(c), and Fig. \ref{fig:s3}-(c).
Then, the joint state space of the MMDP is constructed as $\mathcal{S}=\mathcal{L}^{|\mathcal{N}|}$.
Each agent has at most four actions
- moving forward straight ahead (move forward), forward left, forward right, and stop. 
Some actions will be forbidden at certain states, e.g., moving forward left or right is not allowed if there is only one lane.
We focus on the best-case where agents can always successfully transition among the locations, i.e., transition probability to desired location is one when the vehicle takes desired action.

\begin{table}[h]
\centering
\begin{tabular}{|c|c|c|c|c|}
\hline
\multicolumn{1}{|l|}{\textbf{Scenario ID}} & \multicolumn{1}{l|}{\textbf{Agent ID}} & \multicolumn{1}{l|}{\textbf{DoR}} & \multicolumn{1}{l|}{\textbf{Memory}} & \multicolumn{1}{l|}{\textbf{Runtime}}\\ \hline
\multirow{2}{*}{1}                         & 1                                      & 1   & \multirow{2}{*}{6.69~Mb} & \multirow{2}{*}{3.00 sec}                           \\ \cline{2-3} 
                                           & 2                                      & 0   & &                            \\ \hline
\multirow{3}{*}{2}                         & 1                                      & 0.5  &\multirow{3}{*}{115.29~Mb} &   \multirow{3}{*}{88.49 sec}                             \\ \cline{2-3} 
                                           & 2                                      & 0    & &                               \\ \cline{2-3} 
                                           & 3                                      & 0.5  & &                                 \\ \hline
\multirow{2}{*}{3}                         & 1                                      & 1    & \multirow{2}{*}{5.43~Mb} & \multirow{2}{*}{1.55 sec}                              \\ \cline{2-3} 
                                           & 2                                      & 0    & &                               \\ \cline{2-3} 
                                        
                                           \hline
\end{tabular}
\caption{{Values of Degree of Responsibility (DoR) assigned to vehicle agents following our proposed approach for experiments in CARLA for 
scenarios in Fig. \ref{fig:s1}, Fig. \ref{fig:s2}, and Fig. \ref{fig:s3}. The table additionally shows memory usage and runtime to compute the DoRs in each case.} 
}\label{tab:results}
\end{table}

\subsection{Experiment Results}

Table~\ref{tab:results} presents DoR values 
for each vehicle agent in the three road safety violation scenarios shown in Fig.~\ref{fig:s1} to Fig.~\ref{fig:s3} {along with memory usage and computational runtime}. 

Our method assigns DoR of 1 and 0 to Agents 1 and 2 in Scenario 1. 
This is because Agent 1 could have acted to stop the vehicle and avoid colliding with the pedestrian, regardless of Agent 2's action.
For e.g., Agent 1 could have followed a safe policy, yielding path ($0, \text{move forward, } 3, \text{move forward, }6, \text{stop, } 6$) to avoid 
the pedestrian in location 9, even when Agent 2 follows the same path (1, move forward, 4, move forward, 7, move forward, 10) 
as observed in $\rho$.
Therefore, Agent 1 is fully responsible (i.e., DoR $=$ 1), whereas Agent 2 bears no responsibility. 


For Scenario 2, Agents 1 and 3 share responsibility 
while Agent 2 bears no responsibility for the collision. 
An explanation for these DoR values is that Agent 2 can safely cruise through the road segment by following same actions observed in path $\rho$ regardless of actions of Agents 1 and 3.
However, actions of Agents 1 and 3 cannot guarantee that the safety constraint will be satisfied.
Instead, Agents 1 and 3 will need to cooperate to avoid collision.
For e.g., when Agent 3 follows actions observed in $\rho$, safety can be guaranteed when Agent 1 follows the safe policy (reach location 6 and then stop until Agent 3 reaches location 3).
Similarly, Agent 1 can follow the safe policy even when Agent 3 takes the action observed in $\rho_3$.
Thus, Agents 1 and 3 each have DoR $=0.5$.
Our approach asserts that Agent 1 is fully responsible 
in Scenario 3.
Here, Agent 2 does not have alternate actions available in any counterfactual world except observed actions, i.e., cruising on the highway in the current lane.
The insight underpinning our DoR calculation is that Agent 1 (white SUV) could have taken some alternative actions (e.g., merge into rightmost lane, or stop before merging into highway) to avoid colliding with Agent 2 (orange truck). 
As a result, Agent 1 is assigned $DoR=1$ while Agent 2 is assigned $DoR=0$.

\subsection{{Discussion}}

The notion of DoR developed in this paper and the three scenarios above describe the impact and analysis following a single crash. 
In order to investigate a situation such as a cascade collision of vehicles resulting from the abrupt braking of a leading vehicle will necessitate reasoning about and formulating compositional properties of the DoR. 
Our DoR also does not explicitly account for the type of vehicle (e.g., car, truck) that is involved in an accident. 
Such compositional reasoning may also inform assigning levels of responsibility when multiple types of vehicles are involved in an accident.

Our formulation of DoR solely focuses on quantifying responsibility arising from violations of safety due to erroneous decision-making in multi-agent CPS. 
In practice, there are several other factors that can contribute to ascribing degrees of responsibility to agents involved in an accident. 
Examples of these factors include considerations of conventional norms and the letter of the law \cite{rudisill2017hand}. 
Examining the interplay among safety, legal, and conventional norms and integrating these into a unified quantitative score to assign responsibility following a safety violation will lead to a richer and more interpretable variant of DoR. 


    \section{Conclusion}\label{sec:conclusion}

This paper developed a methodology to identify and quantify responsibility to agents in a multi-agent cyber-physical system when errors in agent decision making led to safety violations. 
The design of a principled automated procedure to assign responsibility was informed by a need to reduce reliance on human expertise and resulting cognitive burden associated with current methodologies. 
The responsibility of an individual agent relative to other agents was quantified using a metric called the degree of responsibility (DoR)  
{to provide an auditable interpretation of agent responsibility using counterfactual reasoning}. 
We considered three instances of safety violations from the National Highway Traffic Safety Administration (NHTSA) and performed extensive experiments using representations of the three scenarios in  
the CARLA urban driving simulator. 
%

We hypothesize that our DoR metric 
can also complement explanation-based techniques focused on reasoning about machine learning models \cite{rozemberczki2022shapley}. 
Future work will provide complete proofs of our theoretical results, and expand empirical validation through additional case studies. 




    \bibliographystyle{IEEEtran}
	\bibliography{MyBib}

\end{document}